\documentclass[a4paper,10pt,aps,prd,showpacs,preprintnumbers,twocolumn,nofootinbib,superscriptaddress]{revtex4-2}

\usepackage[colorlinks=true,
            linkcolor=blue,
            citecolor=blue,
            urlcolor=blue]{hyperref}

\def \be {\begin{equation}}
\def \ee {\end{equation}}
\def \dd {\mathrm{d}} 
\def \t {\tilde}
\def \p {\partial}
\def \l {\left}
\def \r {\right}

\def \bs {\boldsymbol}

\newcommand{\e}[1]{_{\rm #1}}

\newcommand{\beq}{\begin{equation}}
\newcommand{\eeq}{\end{equation}}
\newcommand{\bea}{\begin{eqnarray}}
\newcommand{\eea}{\end{eqnarray}}


\newcommand\ees{\end{eqnarray}}
\newcommand\bees{\begin{eqnarray}}

\def\dd{\mathrm{d}}

\usepackage{lettrine}
\usepackage{graphicx}
\usepackage{booktabs}
\usepackage{comment}
\usepackage[usenames,dvipsnames]{xcolor}
\usepackage{amsmath}
\usepackage{url}
\usepackage{gensymb}
\usepackage{soul}
\usepackage[utf8]{inputenc}
\usepackage{orcidlink}
\usepackage[normalem]{ulem}
\usepackage{empheq}

\def\dd{\mathrm{d}}

\begin{document}

\title{Notes on gravitational wave amplitude evolution\\
  beyond null geodsics}

\author{Charles Dalang\orcidlink{0000-0002-7373-6903}}
\email{charles.dalang@ens.fr}

\affiliation{Institut Philippe Meyer, Département de Physique,\\ École Normale Supérieure (ENS), Université PSL, Paris, France  
}


\date{\today }


\begin{abstract} 
In these notes, we extend a formalism for computing the amplitude of gravitational-waves (GWs) beyond null-geodesic propagation, as may arise in alternative theories of gravity. We derive the wavevector divergence governing the amplitude evolution in this more general setting and apply the formalism to GWs with a frequency-independent subluminal group velocity and to massive gravitons. We show that, in both cases, a naive interpretations of the resulting amplitudes can lead to significant apparent distortions. This formalism opens the door for amplitude interactions with extra degrees of freedom in the presence of dispersion.
\end{abstract}

\maketitle

\section{Introduction} 

General Relativity (GR) is arguably one of the most successful theories developed by humankind. Its rich phenomenology, from the bending of light by stars, to the existence of black holes and gravitational waves (GWs) has been well tested \cite{Will:2014kxa}. 
There are expectations that gravitation requires a deeper understanding at both very high and very low energy scales. Indeed, at very low energy scales, the success of GR relies on the existence of elusive matter contents known as dark matter and dark energy \cite{Planck:2018vyg}. In contrast, at very high energies, a successful theory of quantum gravity is still required \cite{Hawking:1970zqf,Kiefer:2004xyv}. In both scenarios, corrections to the dispersion relation of GWs may appear as a clear hint of beyond GR physics \cite{Barausse:2020rsu, deRham:2018red,Baker:2022eiz, Horava:2009uw, Sefiedgar:2010we, Kostelecky:2016kfm, Amelino-Camelia:2000cpa, Magueijo:2001cr, Amelino-Camelia:2002cqb, Horava:2008ih, Vacaru:2010fi,Garattini:2012km}. 

In GR and in the geometric optics limit, GWs follow null geodesics \cite{Isaacson:1968,Misner:1973prb}. These locally describe a simple dispersion relation, where the angular frequency $\omega$ is directly proportional to the amplitude of the wavevector $\omega = c k$, and the proportionality constant is the speed of light $c$. Since electromagnetic waves follow the same geodesics in vacuum, this means that much of the machinery developed for GWs and electromagnetic signals of cosmological origin is the same \cite{Isaacson:1968, Schneider:1992}. \footnote{Note that the access to phase coherent signals and the wave-optics regime for GWs are two important differences between GWs and light. The latter is due to the fact that GWs are accessible at much larger wavelenghts than light.
} However, in alternative theories of gravity where GWs may obey different dispersion relations, this may not necessarily be the case. The prime examples are powerful constraints on the graviton mass using dispersion of a chirping signal, first proposed in Ref.\,\cite{Will:1997bb}. Indeed, if GWs obey a massive dispersion relation, higher frequencies travel faster than lower ones, such that for a chirping signal, the waveform is a little more compact at the observer than in GR. This results in a time delay between different frequencies which differs from the predictions of GR. This test can be generalized to more sophisticated dispersion relations \cite{Mirshekari:2011yq}, with current dispersion tests focusing on the phase of GWs, assuming that the amplitude is unaffected by the propagation \cite{Baka:2025drk, LIGOScientific:2021sio,deRham:2016nuf,LIGOScientific:2017bnn,LIGOScientific:2019fpa,LIGOScientific:2020tif}. The assumption is that any effect on the amplitude should be negligible while an accumulated phase distortion should be much easier to extract. 

However, if the waveform is compressed and energy is conserved over the waveform, something must also happen to the amplitude. The amplitude evolution of waves in generic curved spacetimes has been studied extensively for null geodesics, which is relevant for light and GWs in GR \cite{Schneider:1992bmb}. In the presence of deviations from null geodesics, some of this machinery requires an extension --  the development and application of which is the subject of the present notes. In this work, we study the amplitude evolution equation for GWs which follow timelike geodesics, which are perturbatively close to null geodesics. In particular, we focus on the case of a graviton mass and a frequency-independent subluminal phase velocity. This is particularly relevant if on top of dispersion, there is energy exchange between the field that modifies the dispersion relation and GWs \cite{Dalang:2020eaj,Menadeo:2025hgf}.

These notes are structured as follows. In Sec.\,\ref{sec:Higher_mode_propagation_equation}, we derive the dispersion relation and amplitude evolution equation for higher harmonics starting from a generic parametrization of the equation of motion of alternative theories of gravity. In Sec.\,\ref{sec:Frequency_evolution}, we study the frequency evolution from the source to the observer, by solving the geodesic equation. In Sec.\,\ref{sec:Amplitude_evolution}, we provide a formula for the wavevector divergence, which controls the amplitude evolution equation for generic dispersion relations. We first discuss the case of subluminal GWs with constant group velocity before discussing a nonzero graviton mass. 
Finally, we discuss limitations and future opportunities in Sec.\,\ref{sec:Discussion}. We use the $(-,+,+,+)$ metric signature and set $\hbar = 1 = c$ throughout the article, except where they appear explicitly.

\section{Higher mode propagation equation}\label{sec:Higher_mode_propagation_equation}

We are interested in theories which have an equation of motion for the metric perturbation of the following form
\begin{align}
(g^{\rho \sigma} + A^{\rho \sigma}) \nabla_\rho \nabla_\sigma h_{\mu\nu} - \mu^2 h_{\mu\nu} = 0\,, \label{eq:EoM_start}
\end{align}
and we assume for simplicity that $A^{\rho \sigma} = \alpha\e{T} u^\rho u^\sigma$, where $u^\sigma$ is a 4-velocity field of observers and $\mu$ represents a graviton mass. The contribution from $A^{\rho\sigma}$ can arise for example in Horndeski theories of gravity \cite{Horndeski:1974wa,Baker:2017hug}, where an extra scalar field interacts nonminimally with gravity. The function $\alpha_T$ corresponds to the familiar Bellini-Sawicki `parameter' \citep{Bellini:2014fua} controlling changes to the propagation speed of GWs (where $\alpha_T=0$ implies luminal propagation). We focus here on second and zeroth order covariant derivatives of the metric perturbation, since in the geometric optics limit, first order derivatives affect directly the amplitude evolution equation in a straightforward way, without affecting the dispersion relation. We expand the metric perturbation in spin weighted spherical harmonics with weight $s=-2$ and the left and right mode on a null tetrad
\begin{align}
h_{\mu\nu} & = \frac{1}{R}\sum_{\ell =0}^{\ell\e{max}} \sum_{m=-\ell}^\ell H_{\mu\nu}^{\ell m}\mathrm{e}^{\mathrm{i} m\varphi}  \,, \label{eq:Ansatz}\\
H_{\mu\nu}^{\ell m} & = \l(h^{\ell m}\e{L} m_\mu m_\nu +  h^{\ell m}\e{R} m^*_\mu m^*_\nu \r)Y_{-2}^{\ell m}(\bs{\hat{n}})\,.
\end{align}
Here $\bs{\hat{n}}$ corresponds to the direction of propagation of the wave at the source. We focus here on the propagation of spin-2 modes and ignore that further propagating degrees of freedom may appear in theories that lead to Eq.\,\eqref{eq:EoM_start}.\footnote{These may lead to detectable extra polarizations, which is a direct signature of these extra degrees of freedom \cite{LISACosmologyWorkingGroup:2026zah}. The focus here is on their indirect effects on the propagation of tensor modes, for example, through the coupling $A^{\rho \sigma}$.} Note that we assume that the source is not precessing, as this generates non-trivial mixing of different $m$ modes of a particular scale $\ell$
\cite{Schmidt:2010it}. The null vector $m^\mu$ and its complex conjugate $m^{*\mu}$ satisfy 
\begin{align}
m^\mu m_\mu = 0\,, \quad m^{*\mu} m^*_\mu =0\,, \quad m^\mu m^*_\mu =1\,.
\end{align}
Our goal is to determine the transport properties of the $h\e{L/R}^{\ell m}$ from the source to the observer. The gradient of the phase defines the wavevector $k_\mu \equiv \p_\mu \varphi$. The geodesic followed by a GW in spacetime $x^\mu(\lambda)$, can be parametrized by an affine parameter $\lambda$ such that $k^\mu = \dd x^\mu/ \dd \lambda $. Plugging the ansatz \eqref{eq:Ansatz} into the equation of motion \eqref{eq:EoM_start} and assuming that the phase varies much faster than the amplitude, one finds
\begin{align}
0 = \sum_{\ell m} & e^{\mathrm{i} m \varphi} \Big[ -m^2 (g^{\rho \sigma} + A^{\rho \sigma})  k_\rho k_\sigma    - \mu^2  \nonumber \\
&  + \mathrm{i} m (g^{\rho \sigma} + A^{\rho \sigma}) (2k_\rho \nabla_\sigma + \nabla_\rho k_\sigma) \Big] H_{\mu\nu}^{\ell m} \label{eq:EoM2}
\end{align}
We assume that the tetrad is parallel transported along the wavevector $k^\mu \nabla_\mu m_\nu =0$ and that it further satisfies $u^\mu \nabla_\mu m_\nu = 0$. In this scenario, one can contract Eq.\,\eqref{eq:EoM2} with $m^{*\mu} m^{*\nu}$ (or $m^\mu m^\nu$) to obtain an equation for $h_L^{\ell m}$ ($h_R^{\ell m}$)
\begin{align}
\sum_{\ell=0}^{\ell\e{max}} \sum_{m=-\ell}^\ell e^{\mathrm{i} m \varphi} a_{\ell m}  Y_{-2}^{\ell m}(\bs{\hat{n}})= 0\,,
\end{align}
with
\begin{align}
 a_{\ell m} & \equiv \Big[ - m^2 (g^{\rho \sigma} + A^{\rho \sigma}) k_\rho k_\sigma  - \mu^2 \nonumber \\
& \quad  + \mathrm{i} m (g^{\rho \sigma} + A^{\rho \sigma}) (2k_\rho \nabla_\sigma + \nabla_\rho k_\sigma) \Big] h\e{L}^{\ell m} \label{eq:EoM4}
\end{align}
The equation of motion is a polynomial in $e^{\mathrm{i}\varphi}$, which has up to $2\ell\e{max}$ roots. Since $\varphi(t)$ spans a continuous set, the coefficients must vanish identically for each $m$
\begin{align}
\sum_{\ell =0}^{\ell\e{max}} a_{\ell m} Y_{-2}^{\ell m}(\bs{\hat{n}}) = 0
\end{align}
Multiplying by $(Y_{-2}^{\ell' m'}(\bs{\hat{n}}))^*$ and integrating over $\bs{\hat{n}}$, one can use the completeness relation  of spin weighted spherical harmonics for fixed spin\footnote{Formally, this requires to assume spherical symmetry of the background spacetime centered on the source. At least the $a_{\ell m}$ should depend weakly on $\boldsymbol{n}$.} $s=-2$, to find that 
\begin{align}
a_{\ell m} = 0\,.
\end{align}
The real and imaginary parts of \eqref{eq:EoM4} must vanish. That is
\begin{align}
-\mu^2 & = (g^{\rho \sigma} + A^{\rho \sigma}) k_\rho k_\sigma \,, \label{eq:Dispersion_Relation}\\
0 & = (g^{\rho \sigma} + A^{\rho \sigma}) (2k_\rho \nabla_\sigma + \nabla_\rho k_\sigma) h\e{L}^{\ell m} \,.\label{eq:Amplitude_Evolution}
\end{align}
The first equation is the dispersion relation while the second one is an evolution equation for the amplitudes $h_L^{\ell m}$. In the next two sections, we discuss the implications for the frequencies and the amplitude of GWs.

\section{Frequency evolution}\label{sec:Frequency_evolution}

The dispersion relation Eq.\,\eqref{eq:Dispersion_Relation} is typically expressed in a local inertial frame as
\begin{align}
\omega^2 = \frac{k^2 + \mu^2}{1-\alpha\e{T}}\,.
\end{align}
When $\mu =0$, one finds the phase velocity
\begin{align}
c\e{T} = \frac{\omega}{k} \simeq 1+\frac{\alpha\e{T}}{2}\,, \label{eq:cT}
\end{align}
to linear order in $\alpha\e{T}$. This clarifies why $\alpha\e{T}$ may be referred as the tensor speed excess. In contrast, when $\alpha\e{T} = 0$, one finds instead the massive dispersion relation
\begin{align}
\frac{\omega}{k} \simeq 1 + \frac{\mu^2}{2\omega^2}\,,\label{eq:cmu}
\end{align}
to linear order in $\mu^2/\omega^2$. In that case, the group velocity $v\e{g} = \p \omega / \p k \simeq 1 - \mu^2/(2\omega^2)$ is subluminal. An alternative way to understand the dispersion relation is to take the covariant derivative of Eq.\,\eqref{eq:Dispersion_Relation} and use the fact that the wavevector is the gradient of a phase. One immediately sees that the wavevector fails to be parallel transported along itself. Instead, it is subject to a force
\begin{align}
k^\mu \nabla_\mu k_\nu = -\frac{1}{2} \nabla_\nu (A^{\rho \sigma} k_\rho k_\sigma)\,.
\end{align}
Note that the mass $\mu$ drops out. One can contract this equation with $u^\nu$ and integrate along $\dd \lambda$, after rearranging to show that the observed frequency relates to the source frequency via
\begin{align}
\omega(\lambda_o) = \omega(\lambda_s) + \int_{\lambda_s}^{\lambda_o} \dd \lambda\, \l(k_\nu k^\mu \nabla_\mu u^\nu - \frac{1}{2} \frac{\dd}{\dd \tau}(A_{\rho \sigma} k^\rho k^\sigma) \r) \,.\label{eq:observed_frequency}
\end{align}
The first term in the integral can be interpreted as a Doppler effect from parallel transporting the 4-velocity along $k^\mu$ and contracting with $k_\nu$. It also appears in GR, and vanishes for static observers in Minkowski spacetime. The second term is a friction term which accumulates along the geodesic. This term is responsible for altering the frequencies of GWs with respect to the expectation from GR. Integrating this last term, we get
\begin{align}
& \int_{\lambda_s}^{\lambda_o} \dd \lambda \,\frac{1}{2}\frac{\dd }{\dd \tau} (A_{\rho \sigma} k^\rho k^\sigma) = \int_{\lambda_s}^{\lambda_o} \dd \lambda \,\alpha\e{T} \omega\frac{\dd \omega}{\dd \tau}\,.
\end{align}
Ignoring the first integral in \eqref{eq:observed_frequency}, effectively assuming a Minkowski spacetime, we get that the source and observed frequencies are related by 
\begin{align}
\omega(\lambda_o) = \omega(\lambda_s)  - \int_{\lambda_s}^{\lambda_o} \dd \lambda\, \alpha\e{T} \omega\frac{\dd \omega}{\dd \tau}\,,\label{eq:observed_omega}
\end{align}
where $\lambda_o$ and $\lambda_s$ stand for the affine parameter of the observer and the source, respectively. Alternatively, one can express the source frequency perturbatively in $\alpha\e{T}$ in terms of the observed one
\begin{align}
\omega(\lambda_s) = \omega(\lambda_o) \l( 1+  \int_{\lambda_s}^{\lambda_o} \dd \lambda\, \alpha\e{T} \frac{\dd \omega}{\dd \tau}\r)\,. \label{eq:source_frequency}
\end{align}
This expression will allow us to express the source waveform, which is naturally expressed in terms of the source frequencies, as a function of the observed frequencies. The presence of $\alpha\e{T}$ affects the observed frequencies, in a way which is degenerate with the source frequencies. The graviton mass $\mu$ in contrast, does not. In the next section, we study the effect on the amplitude of an alternative dispersion relation and how the transformation properties of frequencies (Eq.\,\eqref{eq:source_frequency}) are required to understand the result.

\section{Amplitude evolution}\label{sec:Amplitude_evolution}

The bulk of this work is to study the consequences of deviations from null geodesics for the amplitude of the GWs, namely $h\e{L}^{\ell m}$ (or equivalently $h\e{R}^{\ell m}$ ). To this end, we solve the amplitude evolution equation \eqref{eq:Amplitude_Evolution}, where the mass $\mu$ does not appear explicitly
\begin{align}
(2k^\rho \nabla_\rho + \nabla^\rho k_\rho + 2 A^{\rho \sigma} k_\rho \nabla_\sigma + A^{\rho \sigma}\nabla_\rho k_\sigma ) h\e{L}^{\ell m} = 0 \,. \label{eq:Amplitude_Evolution2}
\end{align}
When $\alpha\e{T} =0$, 
the last two terms vanish and we are left $(2 k^\rho \nabla_\rho + \nabla^\rho k_\rho )h\e{L}^{\ell m} = 0$. In this case, when considering a bundle of geodesics originating from the source, one can use the fact that for null geodesics (which also requires that $\mu=0$), $\nabla^\rho k_\rho = 2 \,\dd \ln D/\dd \lambda$, where $D$ is a comoving distance (see discussion in Sec.\,III.C.3 of \cite{Dalang:2019rke} and Sec.\,2.3 of \cite{Fleury:2015hgz}). Combined with the fact that $h\e{L}^{\ell m}$ behaves like a scalar, such that $k^\rho \nabla_\rho h\e{L}^{\ell m} = \dd  h\e{L}^{\ell m}/\dd \lambda$, this results in the conservation law $\dd/\dd \lambda(h\e{L}^{\ell m} D) = 0$, which implies that the GW amplitude decays with the inverse of the comoving distance $D$. An extra factor of $(1+z)$ arises from converting source frame frequencies and masses to detector frame ones in the amplitude $h\e{L}^{\ell m}$, which yields the well-known decay of GW amplitudes with luminosity distance $D\e{L}$. Here, instead, we show that when the dispersion relation is not null as in Eq.\,\eqref{eq:Dispersion_Relation}, the decay of GWs with distance is more subtle. We compute the leading order correction in $\alpha\e{T}$ or in $\mu^2/\omega^2$, which we call `beyond null geodesic' parameters. We do this by solving Eq.\,\eqref{eq:Amplitude_Evolution} for each $h\e{L}^{\ell m}$. It turns out that the trickiest part of this calculation is to generalize the null geodesic result for the divergence of the wavevector, namely, $\nabla^\mu k_\mu$. In the following, we sketch how this can be done to linear order in beyond null geodesic parameters, by generalizing some results that can be found for null geodesics in Chap.\,2 of \cite{Fleury:2015hgz}. To achieve this, one can define a spacelike vector $d^\mu$, which is along the spatial direction of $k^\nu$ as
\begin{align}
d^\mu \propto (g^{\mu \nu} + u^\mu u^\nu) k_\nu\,.
\end{align}
One can normalize it such that
\begin{align}
u^\mu u_\mu = -1\,, \qquad d^\mu d_\mu = 1\,, \qquad d^\mu u_\mu =0\,.
\end{align}
One can expand the wavevector on this basis such that $k^\mu = \omega u^\mu + k d^\mu$. This is enough to relate an infinitesimal change in spatial distance, in the reference frame of the observer to a small change in affine parameter
\begin{align}
\dd \ell \equiv d_\mu \dd x^\mu = d_\mu k^\mu \dd \lambda = k \dd \lambda\,.
\end{align}
One can also define an orthonormal basis of a screen $\{ s_1^\mu, s_2^\mu\}$, which is orthogonal to $d^\mu$ and $u^\mu$ such that
\begin{align}
s^A_\mu s_B^\mu = \delta^A_B\,, \quad s_\mu^A u^\mu =0 \,, \quad s_\mu^A d^\mu =0\,.
\end{align}
One can define a screen projector, which is most simply expressed in terms of these vectors
\begin{align}
S^{\mu\nu} & \equiv g^{\mu\nu} + u^\mu u^\nu - d^\mu d^\nu\\
& = \delta^{AB} s_A^\mu s_B^\nu\,.
\end{align}
Note that this implies $S^{\mu\nu}u_\mu = 0$, $S^{\mu\nu} d_\mu =0$. We also define the $2\times2$ matrix
\begin{align}
\mathcal{S}_{AB} = s_A^\mu s_B^\nu \nabla_\mu k_\nu \,.
\end{align}
It can be shown that tr$\mathcal{S} =\delta^{AB} \mathcal{S}_{AB} = 2 \dd \ln D /\dd \lambda$ for a bundle of null geodesics, which propagate from a vertex located at the source to the observer (see for example \cite{Fleury:2015hgz}). Since we work formally in spherical symmetry, the direction of propagation of the GWs is not affected by $\alpha\e{T}$ or $\mu$ and we can keep the vectors $d^\mu$, $s_A^\mu$ and $u^\mu$ fixed. A weaker assumption is that any projection of their derivatives varies slower than the wavevector, i.e.\,$\nabla_\nu  e^\mu \ll \omega^{-1} \nabla_\nu  k^\mu $, where $e^\mu$ denotes any of $d^\mu$, $s_A^\mu$ or $u^\mu$, which leads to the same result. The wavevector $k^\mu$ can be expressed in terms of a null vector $\bar{k}^\mu = \omega (u^\mu + d^\mu)$ and a perturbation $\delta k^\mu$
\begin{align}
k^\mu & = \bar{k}^\mu + \delta k^\mu\,, \quad \hbox{with} \quad \delta k^\mu = (k-\omega) d^\mu\,.
\end{align}
Quantities with a bar denote quantities constructed for null geodesics, while $\delta$ indicate terms that are linear in the beyond null geodesic perturbation parameter, namely $\alpha\e{T}$ or in $\mu^2/\omega^2$. For null geodesics, the gradient of the wavevector reads \cite{Fleury:2015hgz}
\begin{align}
\nabla_\mu \bar{k}_\nu & = \mathcal{\bar{S}}_{AB}s^A_\mu s^B_\nu - 2 \omega^{-1}S^\rho_{(\mu} \bar{k}_{\nu)} u^\sigma \nabla_\rho \bar{k}_\sigma \\
& \quad + \omega^{-2} \bar{k}_\mu \bar{k}_\nu u^\rho u^\sigma \nabla_\rho \bar{k}_\sigma
\end{align}
This can be shown starting from $\nabla_\mu \bar{k}_\nu = \delta_\mu^\rho \delta_\nu^\sigma \nabla_\rho \bar{k}_\sigma$ with $\delta^\rho_\mu = S^\rho_\mu + d_\mu d^\rho - u_\mu u^\rho$ and using $\bar{k}^\mu \bar{k}_\mu =0 = \bar{k}^\nu \nabla_\nu \bar{k}_\mu$. By tracing on $\mu$ and $\nu$, one finds
\begin{align}
\nabla^\mu \bar{k}_\mu = \hbox{tr} \bar{\mathcal{S}} =2 \dd \ln D /\dd \bar{\lambda}\,, \label{eq:nabla_k_null}
\end{align}
where the second equality is not trivial and follows from a relatively long derivation presented in \cite{Fleury:2015hgz}. Here, instead we make an expansion around null geodesics. One finds
\begin{align}
\nabla^\mu k_\mu & = \hbox{tr}(\bar{\mathcal{S}}) + \hbox{tr}(\delta \mathcal{S}) + \omega^{-1} (d^\rho - u^\rho) k^\sigma \nabla_\sigma k_\rho \nonumber \\
& \quad + \l( \frac{k^\mu k_\mu}{\omega^2} - \l( \frac{k^3}{\omega^3}- \frac{\omega}{k}\r)\r) u^\rho u^\sigma \nabla_\rho \bar{k}_\sigma \label{eq:nabla_k_1}
\end{align}
where $\hbox{tr}(\bar{\mathcal{S}})$ is given in Eq.\,\eqref{eq:nabla_k_null} and $\delta \mathcal{S}_{AB} = s_A^\mu s_B^\nu \nabla_\mu \delta k_\nu$. Eq.\,\eqref{eq:nabla_k_1} trivially reduces to Eq.\,\eqref{eq:nabla_k_null} in the null geodesic limit. One also needs the trace of $\delta \mathcal{S}_{AB}$, which can be obtained by direct calculation
\begin{align}
\hbox{tr}(\delta \mathcal{S}) & = \delta^{AB} s_A^\mu s_B^\nu \nabla_\mu \delta k_\nu \simeq \frac{k-\omega}{\omega}\hbox{tr}(\bar{\mathcal{S}}) 
\end{align}
where the approximate equality holds because we neglected gradients of $u^\mu$, $s_A^\mu$ or $d^\mu$, next to the derivatives of $k^\mu$ contained in $\hbox{tr}(\bar{S})$. The last term in Eq.\,\eqref{eq:nabla_k_1} contains a contribution from the following expression,
\begin{align}
u^\rho u^\sigma \nabla_\rho \bar{k}_\sigma = -\frac{\dd \omega}{\dd \tau}
\end{align}
which relates to the local proper time derivative of the angular frequency. To relate the affine parameter of the null geodesic to the affine parameter of the timelike geodesic in $\hbox{tr}(\bar{\mathcal{S}}) = 2 \dd \ln D/\dd \bar{\lambda}$, we use $\dd \ell = \omega \dd \bar{\lambda} = k\dd \lambda$. We are left with the generic divergence of the wavevector
\begin{align}
\nabla^\mu k_\mu & = 2 \frac{\dd \ln D}{\dd \lambda} + \omega^{-1}(d^\rho - u^\rho)k^\sigma \nabla_\sigma k_\rho \nonumber \\
& \quad - \l(\frac{k^\mu k_\mu}{\omega^2} - \l( \frac{k^3}{\omega^3} - \frac{\omega}{k} \r) \r) \frac{\dd \omega}{\dd \tau} \,. \label{eq:nablamukmu}
\end{align}
This expression holds for a generic dispersion relation which is perturbatively close to null geodesics. In particular, it reduces to the null geodesic result when $k^\mu k_\mu = 0$ and $k^\sigma \nabla_\sigma k_\rho =0$.

\section{Applications}

In the following, we apply the formalism presented in the previous section to two representative cases, namely subluminal GWs with a frequency independent group velocity in Sec.\,\ref{sec:Time-like_geodesics} and a graviton mass in Sec.\,\ref{sec:Massive_geodesics}. 

\subsection{Subluminal Gravitational waves}\label{sec:Time-like_geodesics}

We first apply our findings for $\nabla^\mu k_\mu$  from Eq.\,\eqref{eq:nablamukmu} 
to the case $\alpha\e{T} \neq 0 = \mu$. In this situation, we expect the different wave frequencies to travel at the same speed $\p\omega/\p k < c$ and therefore the waveform to arrive undistorted to the observer, apart from the dilution with luminosity distance. We clarify how this shows up in the amplitude evolution equation. To leading order in $\alpha\e{T}$, one has
\begin{align}
k^\mu k_\mu & = - A_{\mu\nu} \bar{k}^\mu \bar{k}^\nu = -\alpha\e{T} \omega^2 \label{eq:kmukmu_alphaT} \\
k^\sigma \nabla_\sigma k_\rho & = - \frac{1}{2} \nabla_\rho (A_{\mu\nu} \bar{k}^\mu \bar{k}^\nu) = -\frac{1}{2}\nabla_\rho(\alpha\e{T} \omega^2) \label{eq:geodesic_alphaT}
\end{align}
Some algebra allows us to express the second term in Eq.\,\eqref{eq:nablamukmu} as 
\begin{align}
\omega^{-1} (d^\rho - u^\rho)k^\sigma \nabla_\sigma k_\rho & = - \alpha\e{T}(d^\rho - u^\rho)\nabla_\rho \omega \label{eq:second_term}\\
& = 2\alpha\e{T} \frac{\dd \omega}{\dd \tau} - \alpha\e{T} \omega d^\mu d^\nu \nabla_\mu u_\nu\,. \nonumber \\
& \simeq 2\alpha\e{T} \frac{\dd \omega}{\dd \tau}\,.
\end{align}
Combining Eq.\,\eqref{eq:kmukmu_alphaT}, \eqref{eq:geodesic_alphaT}, \eqref{eq:second_term} and \eqref{eq:cT} in \eqref{eq:nablamukmu}, we obtain
\begin{align}
\nabla^\mu k_\mu = 2 \frac{\dd \ln D}{\dd \lambda} + \omega \frac{\alpha\e{T}}{2} [\nabla_\mu u^\mu - \frac{3}{2} d^\mu d^\nu \nabla_\mu u_\nu] + \alpha\e{T} \frac{\dd \omega}{\dd \tau}
\end{align}
We are left with
\begin{align}
\nabla^\mu k_\mu = 2 \frac{\dd \ln D}{\dd \lambda}+ \alpha\e{T} \frac{\dd \omega}{\dd \tau}\label{eq:T1}
\end{align}
The two last terms in Eq.\,\eqref{eq:Amplitude_Evolution2} are simpler to compute. For the fourth one, $A^{\rho \sigma} \nabla_\rho k_\sigma$ which is obviously linear in $\alpha\e{T}$, one can use the null geodesic result for $\nabla_\rho \bar{k}_\sigma$. After some algebra, one finds
\begin{align}
A^{\rho\sigma} \nabla_\rho \bar{k}_\sigma = -\alpha\e{T} \frac{\dd \omega}{\dd \tau}\label{eq:T2}
\end{align}
The third term in the amplitude evolution equation Eq.\,\eqref{eq:Amplitude_Evolution2}
takes the form 
\begin{align}
2 A^{\rho \sigma} k_\rho \nabla_\sigma h\e{L}^{\ell m} = 2 \alpha\e{T} \omega \frac{\dd h\e{L}^{\ell m }}{\dd \tau}\label{eq:T3}
\end{align}
Plugging \eqref{eq:T1}, \eqref{eq:T2} and \eqref{eq:T3} in the amplitude evolution equation \eqref{eq:Amplitude_Evolution2}, we obtain 
\begin{align}
0& = 2 \frac{\dd h\e{L}^{\ell m}}{\dd \lambda} + \l(2\frac{\dd \ln D}{\dd \lambda} + \alpha\e{T} \frac{\dd \omega}{\dd \tau}\r) h\e{L}^{\ell m} \\
& \quad + 2 \alpha\e{T} \omega \frac{\dd h\e{L}^{\ell m }}{\dd \tau} - \alpha\e{T} h\e{L}^{\ell m} \frac{\dd \omega}{\dd \tau}\,,
\end{align}
which simplifies to 
\begin{align}
\frac{\dd ( h\e{L}^{\ell m} D) }{\dd \lambda} =- \alpha\e{T} \omega  D \frac{\dd h\e{L}^{\ell m }}{\dd \tau}\,.\label{eq:amplitude_non-conservation}
\end{align}
When $\alpha\e{T} = 0$, we get the dilution of the amplitude of $h\e{L}^{\ell m}$ with comoving distance. The right hand side of Eq.\eqref{eq:amplitude_non-conservation} depends on how the amplitude evolves with proper time. In practice, the mode amplitudes $h\e{L}^{\ell m}$ are expressed as power laws in frequencies $h\e{L}^{\ell m}\propto \omega^{c_{\ell m}}$. We consider a single power law as an example. We discuss the sum of multiple power laws, which is the relevant case observationally in App.\,\ref{app:sum_power_laws} and which corresponds to a post-Newtonian expansion of the waveform in GR. If $h\e{L}^{\ell m} = H \omega^{c_{\ell m}}$, this implies that 
\begin{align}
\omega \frac{\dd h\e{L}^{\ell m}}{ \dd \tau} = c_{\ell m} h\e{L}^{\ell m} \frac{\dd \omega}{\dd \tau}\,. \label{eq:time_evolution_waveform}
\end{align}
This equation can nicely be written as a non-conservation equation
\begin{align}
\frac{\dd \ln \l( h\e{L}^{\ell m} D \r) }{\dd \lambda} = - c_{\ell m} \alpha\e{T} \frac{\dd \omega}{\dd \tau} \,.
\end{align}
which can be integrated and linearized in $\alpha\e{T}$ to yield
\begin{align}
h\e{L}^{\ell m}(\lambda_o) 
& \simeq  \frac{h\e{L}^{\ell m}(\lambda_s) D_s}{D_o}\l( 1 - c_{\ell m} \alpha\e{T} \int_{\lambda_s}^{\lambda_o} \dd \lambda \frac{\dd \omega}{\dd \tau}  \r) \,, \label{eq:result_amplitude}
\end{align}
where $D_s \equiv D(\lambda_s)$ represents a comoving distance between the source and a spacetime point  $x^\mu(\lambda_s)$ sufficiently far from it and $D_o = D(\lambda_o)$ is the comoving distance between the source and the observer. At this stage, the waveform appears distorted by the presence of $\alpha\e{T}$. This interpretation however, is naïve. This is because the source amplitude also depends on the source frequencies, which are themselves affected by $\alpha\e{T}$. One must express the source amplitude $h\e{L}^{\ell m}(\lambda_s)$ in terms of the observed frequency $\omega_o$ via Eq.\,\eqref{eq:source_frequency} and linearize in $\alpha\e{T}$ to obtain 
\begin{align}
h\e{L}^{\ell m}(\lambda_s;\omega_o) = H \omega_s^{c_{\ell m}} = H \omega_o^{c_{\ell m}} \l( 1 + c_{\ell m }\alpha\e{T} \int_{\lambda_s}^{\lambda_o} \dd \lambda \frac{\dd \omega}{\dd \tau} \r) \,. \label{eq:h_lm_lambda_s}
\end{align}
The term proportional to $\alpha\e{T}$ in Eq.\,\eqref{eq:h_lm_lambda_s}, which comes from integrating the geodesic equation for the frequencies from the source to the observer cancels out with the one in \eqref{eq:result_amplitude}, which comes from solving the amplitude evolution equation. One is therefore left with
\begin{align}
h\e{L}^{\ell m}(\lambda_o) 
& \simeq  \frac{h\e{L}^{\ell m}(\lambda_s; \omega_o) D_s}{D\e{L}}\,. 
\end{align}
Here, the source amplitude $h\e{L}^{\ell m}(\lambda_s; \omega_o)$ is expressed in terms of the observed frequency $\omega_o$. This is the well-known result that a GW propagating with a frequency-independent group velocity arrives undistorted to the observer. The only effect of the propagation is the damping of the amplitude with the luminosity distance to the observer, when the amplitude is expressed in terms of the observed frequency. While this result is expected, the cancellation is non-trivial and gives us confidence that the formalism (and in particular Eq.\,\eqref{eq:nablamukmu}) is correct. The relevant case observationally is a sum of power laws in frequency, which corresponds to a post-Newtonian expansion waveform in GR. We show in App.\,\ref{app:sum_power_laws} that this result can be generalized to a sum of power laws. Having checked that this nontrivial cancellation happens for a frequency-independent group velocity, we can work out the case of a nonzero graviton mass in the next section.

\subsection{Graviton mass}\label{sec:Massive_geodesics}

Having gained confidence that the beyond null geodesics amplitude evolution calculation yields the correct nontrivial cancellation when $\mu=0$, we now focus on the case $\mu \neq 0 = \alpha\e{T}$, which corresponds to a nonzero graviton mass. The dispersion \eqref{eq:Dispersion_Relation} and amplitude evolution equation \eqref{eq:Amplitude_Evolution} read 
\begin{align}
-\mu^2 & = k^\mu k_\mu\,, \label{eq:kmukmu_mass}\\
0 & = (2 k^\mu \nabla_\mu + \nabla^\mu k_\mu ) h\e{L}^{\ell m}\,. 
\end{align}
In this case, the wavevector is parallel transported along itself and satisfies the geodesic equation
\begin{align}
k^\mu\nabla_\mu k_\nu =0\,.
\end{align} 
In particular, in a generic cosmological setting, the frequencies are redshifted according to\footnote{Indeed, when considering the $\alpha\e{T}\neq 0$ case, it appeared clearer to work in Minkowski space, such that $\omega_o = \omega_s + \mathcal{O}(\alpha\e{T})$, as in Eq.\,\eqref{eq:observed_omega}. However, here there is no obstacle to generalize the relation to a generic cosmological setting since $k^\mu \nabla_\mu k^\nu = 0$ also holds for massive geodesics.}
\begin{align}
\omega_o = \frac{\omega_s}{1+z}\,.
\end{align}
To solve the amplitude evolution equation, we require the wavevector divergence, and use \eqref{eq:nablamukmu} with \eqref{eq:kmukmu_mass}. Writing this out explicitly in terms of $\omega$, $k$, $\mu$ and $\dd \lambda$, this yields
\begin{align}
\nabla^\mu k_\mu & = 2 \frac{\dd \ln D}{\dd \lambda} - \frac{\mu^2}{\omega^2} \frac{\dd \omega}{\dd \tau}\,.
\end{align}
It implies that the evolution equation for the amplitude reads
\begin{align}
2 \frac{\dd h\e{L}^{\ell m}}{\dd \lambda} + 2 \frac{\dd \ln D}{\dd \lambda}h\e{L}^{\ell m}- \frac{\mu^2}{\omega^2}\frac{\dd \omega}{\dd \tau} h\e{L}^{\ell m} = 0\,,
\end{align}
which can be rewritten as a non-conservation equation
\begin{align}
\frac{\dd \ln (h\e{L}^{\ell m} D)}{\dd \lambda} = \frac{\mu^2}{2 \omega^2} \frac{\dd \omega}{\dd \tau}\,.
\end{align}
\begin{align}
h\e{L}^{\ell m}(\lambda_o) = \frac{h\e{L}^{\ell m}(\lambda_s) D_s}{D_o} \l( 1+ \int_{\lambda_s}^{\lambda_o} \dd \lambda \frac{\mu^2}{2 \omega^2} \frac{\dd \omega}{\dd \tau}\r)\,.
\end{align}
Expressing the source amplitude in terms of the observed frequencies allows us to write
\begin{align}
h\e{L}^{\ell m}(\lambda_o,t ) = \frac{h\e{L}^{\ell m}(\lambda_s,\omega_o) D_s}{D\e{L}} \l( 1+ \int_{\lambda_s}^{\lambda_o} \dd \lambda \frac{\mu^2}{2 \omega^2} \frac{\dd \omega}{\dd \tau}\r)\,, \label{eq:observed_hlm_graviton_mass}
\end{align}
since $h\e{L}^{\ell m}(\lambda_s) = h\e{L}^{\ell m}(\lambda_s,\omega_o)/(1+z)$.
The interpretation is the following. Since frequencies travel unaffected by the graviton mass, i.e.\,in Minkowski spacetime $\omega_o= \omega_s$ and higher frequencies travel faster than lower ones, the end of a chirping signal catches up the beginning of the waveform. This implies that a finite-duration waveform is eventually more compact at the observer than it was at the source. Since the energy of the waveform should be conserved over the volume spanned by the area $4 \pi D\e{L}^2$ times the length spanned by the waveform, this requires the amplitude of the wave to be shifted up. The amount by which this happens is the postfactor that we have calculated in Eq.\,\eqref{eq:observed_hlm_graviton_mass}. Let us express the correction to the amplitude,
\begin{align}
\int_{\lambda_s}^{\lambda_o} \dd \lambda \frac{\mu^2}{\omega^2}\frac{\dd \omega}{\dd \tau} = A \mu^2  D_o \omega_o^{2/3}
\end{align}
since  $\dd \omega/\dd \tau = A \omega^{11/3}$ to lowest Post-Newtonian order \cite{Maggiore:2007ulw}. Therefore, the observed waveform amplitude can be rescaled according to 
\begin{align}
h\e{L}^{\ell m}(\lambda_o) = \frac{h\e{L}^{\ell m}(\lambda_s;\omega_o) D_s}{D\e{L}} \l( 1+ \mu^2 B \omega_o^{2/3} \r)\,. \label{eq:observed_hlm_Bm}
\end{align}
where, reintroducing units of $\hbar$ and $c$, we defined
\begin{align}
B \equiv \frac{
6\cdot 2^{1/3}}{5\hbar^2 c^2}  \l( G \mathcal{M}_z\r)^{5/3} D_o\,,
\end{align}
where $\mathcal{M}_z$ indicates the redshifted chirp mass. This amplitude shift is common to all higher harmonics. Its frequency dependence suggests that it is not degenerate with the luminosity distance and that it should be accounted for in tests of the dispersion relation. However, this time-domain interpretation is incomplete. As it turns out, this frequency-dependent amplitude evolution disappears in the frequency domain. This can be understood as follows. The time-domain amplitude can be calculated assuming that the frequency domain amplitude is unaffected. In the stationary phase approximation, the frequency domain amplitude reads \cite{Hu:2024wub}
\begin{align}
\t{h}(f_o) = h\e{MG}(t) \sqrt{\frac{\p t\e{MG}}{\p f_o}} = h\e{GR}(t) \sqrt{\frac{\p t\e{GR}}{\p f_o}}\,, \label{eq:conserved_hf}
\end{align}
where $f_o= \omega_o/(2\pi)$, $\t{h}(f)$ indicates the Fourier transform of the GW amplitude, $h\e{MG}$ and $h\e{GR}$ indicate the time-domain amplitude for the modified gravity with a graviton mass and its general relativistic counterpart, respectively. The arrival time of each frequency to lowest PN order in GR is given by \cite{Hu:2024wub}
\begin{align}
t\e{GR}(f_o) = t_c - 5 (4\pi f_o)^{-8/3} (G\mathcal{M}_z)^{-5/3}\,,\label{eq:t_GR}
\end{align}
where $t_c$ is the coalescence time. In the presence of a massive graviton, each frequency travels slower than its GR counterpart and arrives a little bit later \cite{Colangeli:2025hrs}
\begin{align}
t\e{MG}(f_o) = t\e{GR}(f_o) + \frac{\mu^2}{8\pi^2 f_o^2} D_o\,.\label{eq:t_MG}
\end{align}
Taking the derivatives of Eq.\,\eqref{eq:t_GR} and \eqref{eq:t_MG} with respect to frequency also leads Eq.\,\eqref{eq:conserved_hf} to 
\begin{align}
h\e{MG}(t) = h\e{GR}(t) \l( 1+ \mu^2 B \omega_o^{2/3} \r)\,,\label{eq:result_conserved_hf}
\end{align}
which agrees with Eq.\eqref{eq:observed_hlm_Bm}. Eq.\,\eqref{eq:result_conserved_hf} was obtained assuming that the frequency domain amplitude is unaffected, while Eq.\,\eqref{eq:observed_hlm_Bm} was obtained by direct calculation of the evolution of the amplitude. Hence, we conclude that we have provided a formalism which allows to directly compute the time domain waveform amplitude, without assuming \textit{à priori} that the frequency domain waveform amplitude is conserved. While for a graviton mass, the amplitude in each frequency is conserved, this need not be the case for more complicated dispersion relations in the presence of interactions, as may be the case in Horndeski theories \cite{Menadeo:2025hgf}. In this case, the formalism that we provide allows to directly compute corrections to the amplitude.

\section{Discussion}\label{sec:Discussion}

In these notes, we have explored a formalism to directly compute the amplitude of GWs beyond null geodesics. We derived an equation that rules the amplitude evolution of GWs, which allows for deviations from null geodesics in the form $k^\mu k_\mu \neq 0$ and when the wavevector is subject to a force ($k^\mu \nabla_\mu k^\nu \neq 0$). We have shown that for time-like geodesics with a subluminal speed of GWs, there are subtle cancellations between the transport of the amplitude and frequencies between the source and the observer. If those are not correctly accounted for, this may lead to a spurious frequency-dependent modulation of the amplitude. However, consistently taking into account the transport of the amplitude and of the frequency leads to no modulation of the amplitude, which is only diluted by the luminosity distance between the source and the observer. 

We have then applied the formalism to massive gravitons. By direct calculation, we found that the time-domain waveform amplitude is shifted up for a chirping signal. While this is necessary for energy conservation, we have shown how this time-domain shift results in a frequency domain amplitude which is unaffected by the graviton mass. This implies that tests of the graviton mass using the dispersion relation can be applied consistently by only taking into account modifications of the phase. 

In the presence of extra fields, interactions may imply that energy leaks from one field to another. This was shown not to be the case for Horndeski theories in which tensor waves propagate at the speed of light \cite{Dalang:2020eaj}. However, when those obey a more sophisticated dispersion relation, interactions between scalar waves and tensor waves become possible \cite{Menadeo:2025hgf}. The formalism that we propose is ideally suited for this purpose.

\vspace{5mm}

\textbf{Acknowledgements} It is a pleasure to thank Pierre Fleury, Serena Giardino, Nicola Menadeo, Miguel Zumalac\'arregui, Emeric Seraille, Sebastian Von Hausegger, Tom Bertheas, Alice Albouy, Cédrid Deffayet, Claudia de Rham, Elena Colangeli, Luc Blanchet for helpful discussions. I would also like to  Michael Williams and Tessa Baker for useful discussions and comments on a preliminary version of this work. I would also like to thank Nathan Johnson-McDaniel for very helpful comments on a preliminary version of this work.

\begin{appendix}
\section{Effects on the amplitude of timelike geodesics for a more realistic waveform 
}\label{app:sum_power_laws}

In Sec.\,\ref{sec:Time-like_geodesics}, we computed the transport properties of the waveform amplitude by assuming that it is a perfect power law in frequency, i.e.\,$h\e{L}^{\ell m} \propto \omega^{c_{\ell m}}$. This simple Ansatz allowed us to proceed and to show that the effect of $\alpha\e{T}$ on the amplitude cancels with an analogous effect on the transport of the frequency. However, in practice, the waveform higher modes are usually expressed as a PN expansion \cite{Blanchet:2013haa}. The expansion may be written as a sum of powerlaws in frequency
\begin{align}
h\e{L}^{\ell m} = \sum_{k=1}^N a^{\ell m}_k \omega^{n_k}\,,
\end{align}
where $a^{\ell m}_k$ are PN coefficients. For example, the coefficients and powers have been calculated for quasi-circular inspiral waveforms without tidal effects \cite{Blanchet:2023bwj}. One may wonder whether the cancellation is exact in this case. In the following, we show that the cancellation also holds in this situation. The proper time derivative of the amplitude on the left hand side of \eqref{eq:time_evolution_waveform} rather reads
\begin{align}
\omega \frac{\dd h\e{L}^{\ell m}}{\dd \tau } = \frac{\dd \omega}{\dd \tau}  \underbrace{\sum_{k=1}^N a^{\ell m}_k n_k \omega^{n_k} }_{\neq h\e{L}^{\ell m}}
\end{align}
It is quite important to appreciate that this term is not proportional to $h\e{L}^{\ell m}$. The proportionality is broken by the presence of multiple powerlaws with different indices $n_k$, which evolve differently with proper time.
We solve Eq.\,\eqref{eq:amplitude_non-conservation} perturbatively in $\alpha\e{T}$ such that 
\begin{align}
\frac{\dd \ln h\e{L}^{\ell m} D}{\dd \lambda} = -\alpha\e{T} \frac{\sum_{k=1}^N n_k a^{\ell m}_k  \omega^{n_k}}{\sum_{j=1}^N a^{\ell m}_j  \omega^{n_j}} \frac{\dd \omega}{\dd \tau} 
\end{align}
We obtain
\begin{align}
 h\e{L}^{\ell m}(\lambda_o) & = \frac{ h\e{L}^{\ell m}(\lambda_s) D_s}{D_o} \\
& \quad \times \l( 1 -\alpha\e{T} \int_{\lambda_s}^{\lambda_o} \dd \lambda\, \frac{\sum_{k=1}^N n_k a^{\ell m}_k  \omega^{n_k}}{\sum_{j=1}^N a^{\ell m}_j  \omega^{n_j}} \frac{\dd \omega}{\dd \tau} \r) \nonumber 
\end{align}
This time, expressing the amplitude in terms of the observed frequency is also different
\begin{align}
h\e{L}^{\ell m}(\lambda_s) & = \sum_{k=1}^N a_k^{\ell m} \omega_s^{n_k} \\
& = \sum_{k=1}^N a_k^{\ell m} \omega_o^{n_k} \l( 1+ n_k \alpha\e{T}   \int_{\lambda_s}^{\lambda_o} \dd \lambda\,\frac{\dd \omega}{\dd \tau} \r)
\end{align}
We are left with
\begin{align}
h\e{L}^{\ell m}(\lambda_o) & = \frac{D_s}{D_o} \sum_{k=1}^N a_k^{\ell m} \omega_o^{n_k} \l( 1+ n_k \alpha\e{T}   \int_{\lambda_s}^{\lambda_o} \dd \lambda\,\frac{\dd \omega}{\dd \tau} \r) \nonumber \\
& \quad \times \l( 1 -\alpha\e{T} \int_{\lambda_s}^{\lambda_o} \dd \lambda\, \frac{\sum_{k=1}^N n_k a^{\ell m}_k  \omega^{n_k}}{\sum_{j=1}^N a^{\ell m}_j  \omega^{n_j}} \frac{\dd \omega}{\dd \tau} \r)\label{eq:h_lm_amplitude}
\end{align}
One can compute the integral in Eq.\,\eqref{eq:h_lm_amplitude}, for a chirping signal $\dd \omega/\dd \tau = A \omega^{11/3}$. Using $\dd \lambda = \dd D/\omega$ and the fact that to zeroth order in $\alpha\e{T}$, $\omega_o =\omega_s$, we get
\begin{align}
\alpha\e{T}\int_{\lambda_s}^{\lambda_o} \dd \lambda \frac{\dd \omega}{ \dd \tau} = \alpha\e{T} A \omega_o^{8/3} D_o
\end{align}
where 
\begin{align}
A = \frac{12 \cdot 2^{1/3}}{5} \l( \frac{G \mathcal{M}_z}{c^3}\r)^{5/3}\,.
\end{align}
Likewise
\begin{align}
\int_{\lambda_s}^{\lambda_o} \dd \lambda\, \frac{\sum_{k=1}^N n_k a^{\ell m}_k  \omega^{n_k}}{\sum_{j=1}^N a^{\ell m}_j  \omega^{n_j}} \frac{\dd \omega}{\dd \tau} & = A \omega_o^{8/3} D_o \nonumber  \\
& \quad \times \frac{\sum_{k=1}^N n_k a^{\ell m}_k  \omega_o^{n_k}}{\sum_{j=1}^N a^{\ell m}_j  \omega_o^{n_j}} 
\end{align}
Therefore, the amplitudes of the higher harmonics at the observer read
\begin{align}
h\e{L}^{\ell m}(\lambda_o) & = \frac{D_s}{D_o} \l( 1 -\frac{\alpha\e{T} A D_o \omega_o^{8/3}}{c}  \frac{\sum_{i=1}^N n_i a^{\ell m}_i  \omega_o^{n_i}}{\sum_{j=1}^N a^{\ell m}_j  \omega_o^{n_j}} \r) \nonumber\\
& \quad \times  \sum_{k=1}^N a_k^{\ell m} \omega_o^{n_k} \l( 1+ \frac{n_k \alpha\e{T}  A D_o \omega_o^{8/3}}{c} \r) \\
& = \frac{D_s}{D_o} \l( 1 -\frac{\alpha\e{T} A D_o \omega_o^{8/3}}{c}  \frac{\sum_{i=1}^N n_i a^{\ell m}_i  \omega_o^{n_i}}{\sum_{j=1}^N a^{\ell m}_j  \omega_o^{n_j}} \r) \nonumber\\
& \quad \times  \sum_{k=1}^N a_k^{\ell m} \omega_o^{n_k} \l( 1+ \frac{\alpha\e{T}  A D_o \omega_o^{8/3}}{c} \frac{\sum_{i=1}^N n_i a_i^{\ell m}\omega_o^{n_i}}{\sum_{j=1}^N a_j^{\ell m} \omega_o^{n_j}}\r) \nonumber \\
& =\frac{D_s}{D_o} \sum_{k}^{\ell m}a_k^{\ell m}\omega_o^{n_k}  = \frac{D_s}{D_o}h\e{L}^{\ell m}(\lambda_s;\omega_o)\,.
\end{align}
Therefore, the amplitude is only affected by the dilution with distance, as in the simple power law case discussed in Sec.\,\ref{sec:Time-like_geodesics}. It turns out that considering a more generic waveform described by a PN expansion leaves the perfect cancellation between amplitude and frequency intact. This should come as no surprise since we are computing what happens to the amplitude of a waveform for which the dispersion relation is trivial $\omega= c\e{T} k$. Nonetheless, it may be satisfying to see this cancellation happening explicitly here.

\end{appendix}

\bibliographystyle{apsrev4-2}
\bibliography{references} 

\end{document}